


\input psfig 
\input phyzzx
\font\first=cmb10 scaled\magstep3
\font\second=cmr10 scaled\magstep2
{\nopagenumbers
\rightline{Preprint RIMS 965}
\rightline{January 1994}
\rightline{hep-th/9402051}
\vskip2cm
\centerline{\first Extended reflection equation algebras, the braid
group on}
\smallskip
\centerline{\first  a handlebody and associated link polynomials}
\vskip2cm
\centerline{{{\second Christian Schwiebert}\footnote{\star}
{Supported by the Science and Technology Fellowship Programme for
Japan under the auspices of the Commission of the European
Communities}}\footnote{\diamondsuit}{e-mail:
{\tt cbs@kurims.kyoto-u.ac.jp}}}
\centerline{\it Research Institute for Mathematical Sciences}
\centerline{\it Kyoto University, Sakyo-ku, 606 Kyoto, Japan}
\vskip3cm
\centerline{ABSTRACT}
\smallskip
\noindent
The correspondence of the braid group on a handlebody of arbitrary
genus to the algebra of Yang-Baxter and extended reflection equation
operators is shown. Representations of the infinite dimensional
extended reflection equation algebra in terms of direct products of
quantum algebra generators are derived, they lead to a representation
of this braid group in terms of $R$-matrices. Restriction to the
reflection equation operators only gives the coloured braid group.
The reflection equation operators, describing the effect of handles
attached to a 3-ball, satisfy characteristic equations which give rise
to additional skein relations and thereby invariants of links on
handlebodies. The origin of the skein relations is explained and they
are derived from an adequately adapted handlebody version of the Jones
polynomial. Relevance of these results to the construction of link
polynomials on closed 3-manifolds via Heegard splitting and surgery is
indicated.
\vfill
\break}
\pageno=2


\def \ADW {Y. Akutsu, T. Deguchi and M. Wadati,
J. Phys. Soc. Jpn. {\bf 56} (1987) 839, 3039, 3464;
{\bf 57} (1988) 757, 1173, 1905}
\def \Ale {A. Yu. Alekseev, Uppsala Univ. preprint, hep-th/9311074}
\def \AFS {A. Yu. Alekseev, L. D. Faddeev and M. A.
Semenov-Tian-Shansky, Commun. Math. Phys. {\bf 149} (1992) 335}
\def \Alex {J. W. Alexander, Trans. Am. Math. Soc. {\bf 20} (1923) 275;
Proc. Natl. Acad. Sci. {\bf 9} (1928) 93}
\def \AGS {L. Alvarez-Gaume, C. Gomez and G. Sierra, Nucl. Phys. {\bf
B330} (1990) 347}
\def \Art {E. Artin, Ann. Math. {\bf 48} (1947) 101}
\def \Che {I. V. Cherednik, Teor. Mat. Fiz. {\bf 61} (1984) 55}
\def \CDSWZ {C. Chryssomalakos et. al., Commun. Math. Phys. {\bf 147}
(1992) 634}
\def \CWSSW {U. Carow-Watamura et. al., Z. Phys. {\bf C48} (1990) 159}
\def \Con {J. H. Conway, in \lq\lq Computational problems in abstract
algebra \rq\rq, Pergamon, New York 1970}
\def \FRT {L. D. Faddeev, N. Yu. Reshetikhin and L. A. Takhtajan,
Alg. i Anal. {\bf 1} (1989) 178 (in Russian, English transl.:
Leningrad Math. J. {\bf 1} (1990) 193)}
\def \GMM {E. Guadagnini, M. Martellini and M. Mintchev, \ in: \lq\lq
Proceedings of the ${\rm 13^{th}}$ John Hopkins workshop on current
problems in particle theory\rq\rq, ed. L. Lusanna, World Scientific,
Singapore 1990}
\def \Guaa {E. Guadagnini, \ in: \lq\lq Proceedings of the
${\rm 14^{th}}$ John Hopkins workshop on current problems in particle
theory\rq\rq, eds. G. Domokos et. al., World Scientific, Singapore 1991}
\def \Guab {E. Guadagnini, Nucl. Phys. {\bf B375} (1992) 381}
\def \Guac {E. Guadagnini and S. Panicucci, Nucl. Phys. {\bf B388}
(1992) 159}
\def \Jon {V. F. R. Jones, Bull. Am. Math. Soc. {\bf 12} (1985) 103;
Ann. Math. {\bf 126} (1987) 335}
\def \Jur {B. Jur{\v c}o, Lett. Math. Phys. {\bf 27} (1991) 177}
\def \Kauf {L. H. Kauffman, Topology {\bf 26} (1987) 395}
\def \KaZa {M. Karowski and A. Zapletal, FU Berlin preprint,
hep-th/9312008}
\def \Koha {\lq\lq New developments in the theory of knots\rq\rq,
ed. T. Kohno, World Scientific, Singapore 1989}
\def \Kohb {T. Kohno, Topology {\bf 31} (1992) 203}
\def \KuSa {P. P. Kulish and R. Sasaki, Durham Univ. preprint DTP-92-53}
\def \KSS {P. P. Kulish, R. Sasaki and C. Schwiebert, J. Math. Phys.
{\bf 34} (1993) 286}
\def \Kula {P. P. Kulish, \  in: \lq\lq Proceedings of the second
international Wigner symposium\rq\rq,  eds. H. D. Doebner et. al.,
World Scientific, Singapore 1993}
\def \Kulb {P. P. Kulish, Kyoto Univ. preprint YITP/K-92-984}
\def \Mada {S. Majid, J. Math. Phys. {\bf 32} (1991) 3246}
\def \Madb {S. Majid, J. Math. Phys. {\bf 34} (1993) 1176}
\def \Man {Yu. I. Manin, Montreal Univ. preprint CRM-1561 (1988)}
\def \MoRe {G. Moore and N. Yu. Reshetikhin, Nucl. Phys. {\bf B328}
(1989) 557}
\def \Pod {P. Podles, Lett. Math. Phys. {\bf 14} (1987) 193}
\def \ReTu {N. Yu. Reshetikhin and V. G. Turaev, Commun. Math. Phys.
{\bf 127} (1991) 1; Invent. Math. {\bf 103} (1991) 547}
\def \Sch {C. Schwiebert, Kyoto Univ. preprint YITP/U-92-38, unpublished}
\def \Sos {A. B. Sossinsky, \  in:
\lq\lq  Euler international mathematical institute on quantum
groups\rq\rq,  ed. P. P. Kulish, Lect. Notes Math. {\bf 1510},
Springer, Berlin 1992}
\def \SmWZ {W. B. Schmidke, J. Wess and B. Zumino, Z. Phys. {\bf C52}
(1991) 471}
\def \Skla {E. K. Sklyanin, J. Phys. {\bf A21} (1988) 2375}
\def \Sklb {P. P. Kulish and E. K. Sklyanin, J. Phys. {\bf A25} (1992)
5963}
\def \SWZ {P. Schupp, P. Watts and B. Zumino, Commun. Math. Phys.
{\bf 157} (1993) 305}
\def \Tak {L. A. Takhtajan, \ in: \lq\lq Introduction to quantum
groups and integrable massive models of quantum field theory\rq\rq,
eds. M. L. Ge and B. H. Zhao, World Scientific, Singapore 1991}
\def \Tur {V. G. Turaev, Invent. Math. {\bf 92} (1988) 527}
\def \Wita {E. Witten, Commun. Math. Phys. {\bf 121} (1989) 351}
\def \Witb {E. Witten, Nucl. Phys. {\bf B322} (1989) 629}
\def \WeZu {J. Wess and B. Zumino, Nucl. Phys. (Proc. Suppl.)
{\bf B18} (1990) 302}
\def \Zum {B. Zumino, in: \lq\lq Proceedings of the ${\rm X^{th}}$
congress on mathematical physics\rq\rq, ed. K. Schm\"udgen, Springer,
Berlin 1992}


\def\sl#1{$sl_q(#1)$}
\def\Rt{\widetilde R}
\def\Ri{R^{-1}}
\def\Rh{\widehat R}
\def\Rth{\widehat {\widetilde R}}
\def\k#1#2{K^{#1}_{(#2)}}
\def\l#1#2{L^{#1}_{#2}}
\def\h#1#2{S^{#1}_{#2}}
\def\rea{RE algebra}
\def\a{\alpha}
\def\b{\beta}

\def\d{\delta}
\def\D{\Delta}
\def\e{\varepsilon}
\def\s{\sigma}
\def\t{\tau}
\def\w{\omega}
\def\NPrefs{\let\refmark=\NPrefmark}
\NPrefs


\chapter{INTRODUCTION}

In this paper\footnote\dag{A completely rewritten and largely
extended version of ref. \lbrack 10\rbrack} we should like to explain
the relation between quantum groups (QG) and the braid group on a
three dimensional manifold of arbitrary genus with boundary. As a
consequence we will be able to define invariants of links on such
manifolds. A three-manifold having a genus $g$ Riemann surface as
boundary is conventionally named a genus $g$ handlebody. The braid
group on such a handlebody can be formulated in terms of the usual
braid group generators $\s_i$ for genus zero 3-manifolds
\Ref\rArt{\Art} plus additional generators $\t_\a$ implementing
windings around handles. We shall make use of the results in
\Ref\rSos{\Sos} where such a description was given, explicitly, the
braid group $B^g_n$ on a handlebody comprises the following relations:
$$\eqalign{\s_i \s_{i+1} \s_i &= \s_{i+1} \s_i \s_{i+1} ,
           \quad i=1,\ldots,n-1 \cr
\s_i \s_j &= \s_j \s_i , \quad \mid i-j \mid \geq 2 \cr
\s_i \t_\a &= \t_\a \s_i , \quad i \geq 2 , \ \a =1,\ldots,g \cr
\s_1 \t_\a \s_1 \t_\a &= \t_\a \s_1 \t_\a \s_1 , \cr
\s_1 \t_\a \s_1^{-1} \t_\b &= \t_\b \s_1 \t_\a \s_1^{-1} ,
\quad  \a < \b .} \eqn \bralgf $$
The first two equations define the well known Artin braid group $B_n$
acting on $n$ strands in a topologically trivial 3-manifold
\refmark{\rArt}. For each handle there is a new generator $\t_\a$
having nontrivial commutation relations only with $\s_1$ and the $\t$
generators. The last equation is absent in the case of a solid torus,
i.e. for genus one. The first equation of \bralgf\ corresponds to the
Yang-Baxter equation written in braid form
$$ \Rh_{12} \Rh_{23} \Rh_{12} = \Rh_{23} \Rh_{12} \Rh_{23} ,
\eqn \ybef $$
providing a link to the theory of quantum groups as $\s_i$ can be
represented in terms of the $\Rh$-matrix. The fourth equation of
\bralgf\ is also related to quantum groups, it can be considered
both as a comodule invariant w.r.t. QG coaction and as a way of
describing the quantum algebra, we will use it in the form
$$ R_{12} K_1 R_{21} K_2 = K_2 R_{12} K_1 R_{21} . \eqn \ref $$
There exist also spectral parameter dependent versions of it which play
prominent roles in quantum inverse scattering \Ref\rSkla{\Skla},
describing the commutation relations of monodromy matrices. Actually,
they appeared first in the study of two particle scattering on a
half-line, with matrix $K(\theta)$ describing reflection of a particle
at the endpoint and $R(\theta - {\theta}')$ describing two particle
scattering \Ref\rChe{\Che}. Hence the name reflection equation (RE),
suggested in \Ref\rKula{\Kula}, where also the connection of \ref\
with $B^1_n$ was mentioned. We should mention that there exists still
another spectral parameter independent reflection equation
\REF\rKSS{\KSS} \REF\rSklb{\Sklb} \refmark{\rKSS,\rSklb}, which is
invariant w.r.t. different QG comodule transformations compared to
\ref. The last equation of \bralgf\ is close to the RE, it is a
compatibility condition for solutions of the RE s.t. these can be
combined into new solutions of the RE. In QG language it looks like
$$  R_{12}^{\phantom{-1}} K^{\phantom{1}}_1 R^{-1}_{12}
{K'}^{\phantom{1}}_2 = {K'}^{\phantom{1}}_2 R^{\phantom{-1}}_{12}
K^{\phantom{1}}_1 R^{-1}_{12} ,  \eqn \retwif  $$
and its properties and connection to $B^g_n$ were
discussed in \REF\rKulb{\Kulb} \REF\rKuSa{\KuSa} \REF\rSch{\Sch}
\refmark{\rKulb,\rKuSa,\rSch} under different aspects.
When discussing representations of the braid group \bralgf\ we will
naturally be led to representations of $\t_\a$ in terms of
$R$-matrices s.t. $B^g_n$ can be viewed as a subgroup of $B_{n+g}$.
Equivalently, $\t_\a$ can be expressed in terms of quantum algebra
generators. Indeed we will derive whole series of new solutions of both
\ref\ and \retwif\ in terms of quantum algebra generators, and they
precisely correspond to the description of $\t_\a$ in terms of
$R$-matrices. Furthermore, we derive quadratic characteristic
equations for the matrix $K$, and hence the additional generators
$\t_\a$, similar to the Hecke algebra relation for $\s_i$. They can be
interpreted as an additional skein relation when considering closed
braids on the handlebody and, in principle, they recursively define
link invariants for closed braids on arbitrary genus 3-manifolds with
boundary \refmark{\rSch}. This in turn, via Heegaard splitting, might
be a way of constructing invariant polynomials of links on arbitrary
3-manifolds without boundary. Since we know the representation of
$\t_\a$ in terms of $R$-matrices we can also write down a trace
formula for the link invariants, this is equivalent to using the
quantum trace that is defined for the matrix $K$ \refmark{\rKSS}.
Further we show that the characteristic equation for $\t_\a$ is
actually a consequence of the one for $\s_i$.

The plan of the paper is as follows. We introduce the RE in section
two and discuss its properties as an associative quadratic algebra,
then we extend it by \retwif\ and derive new solutions of the combined
system in terms of quantum algebra generators. In section three we
review some results of \refmark{\rSos} concerning the braid group on a
handlebody and obtain the representation of $\t_\a$ in terms of
$R$-matrices and quantum algebra generators. We also discuss there the
connection between the Hecke algebra relation and the quadratic
equation for $\t_\a$.  In the fourth section we look at closed braids
on handlebodies and their invariants, notably by means of new skein
relations and quantum traces for the additional generators. Finally,
in the fifth section we mention some implications and possible
applications of our results.

\break


\chapter{ALGEBRAS OF REFLECTION EQUATION OPERATORS}

We will study the properties of the following reflection
equation\footnote\dag{We assume familiarity of the reader with
basic quantum group terminology as introduced in \REF\rFRT{\FRT}
\REF\rTak{\Tak} \refmark{\rFRT,\rTak}, for example. Throughout this
paper when giving explicit examples we only use \sl2 for simplicity,
generalizations should be obvious.}
$$ R K_1 \Rt K_2 = K_2 R K_1 \Rt , \eqn \re $$
where $\Rt = P R P$ and $P$ is the permutation operator. Its basic
property and a guideline for its construction is invariance w.r.t. the
QG coaction, i.e.  $K_T = T K T^{-1}$ is also a solution of this RE if
all elements of $K$ and $T$ commute, $\lbrack K_{ij},T_{mn} \rbrack =
0$, and $T$ obeys the QG relations
$$ R T_1 T_2 = T_2 T_1 R . \eqn \rtt  $$
Just as in the case of the defining relations \rtt\ of the QG we can
view \re\  as an associative quadratic algebra. If we use the \sl2
$R$-matrix
$$ {R^{ij}}_{kl} = \pmatrix{q&0&0&0 \cr 0&1&0&0 \cr 0&\w&1&0 \cr
0&0&0&q \cr} , \quad \w=q-q\sp {-1}  \eqn\rmat $$
being a solution of the Yang-Baxter equation
$$ R_{12} R_{13} R_{23} = R_{23} R_{13} R_{12} ,  \eqn \ybe $$
then we find that the commutation relations for the
entries of the matrix {\hbox{$K =$  $a\,b\choose c\,d$}}
are given by
$$ \eqalign{ab&=q\sp {-2}ba , \cr  ac&=q\sp 2ca , \cr}  \qquad
\eqalign{ad&=da , \cr  bc-cb&=q\sp {-1}\w (ad-a\sp 2) , \cr}  \qquad
\eqalign{bd-db&=-q\sp {-1}\w ab , \cr  cd-dc&=q\sp {-1}\w ca  .
\cr} \eqn \alg  $$
This algebra has two central elements, the quantum trace and the
quantum determinant which we set equal to one
$$ c_1 = q^{-1} a + q d , \qquad  c_2 = a d - q^2 c b \equiv 1 . \eqn
\cent $$
The normalization of $c_1$ is chosen such that $K$ and $K^{-1}$ have
equal quantum trace. Using these relations the `antipode' $S(K)\equiv
K^{-1}$ can be found to be
$$ K^{-1} = \pmatrix{q^2 d - q \w a & -q^2 b \cr
                     -q^2 c & a \cr} . \eqn \inv $$
Then we easily establish a relation (characteristic equation) between
$K$ and $K^{-1}$
$$ q K + q^{-1} K^{-1} - c_1 I = 0 , \eqn \cheq  $$
which will give rise to a skein relation later on.

\noindent
A few remarks about the properties of the above algebra follow.
\nextline
{\it (i)\phantom{vii}} The RE algebra \alg\ depends only on $q^2$.
\nextline
{\it (ii)\phantom{vi}} If $q$ is a root of unity, $q^{2p}=1$, then
$a^p$ is a further central element. \nextline
{\it (iii)\phantom{v}} The \sl2 \rea\  has two constant or
one-dimensional representations, one of them clearly is the identity
matrix and the other one a lower-right triangular matrix with
arbitrary constants $b,c,d$. Constant solutions of \re\ were studied
in \refmark{\rKSS}. \nextline
{\it (iv)\phantom{ii}} The $K$-matrix can be considered as a
product of two suitable quantum planes \REF\rMan{\Man}
\REF\rWeZu{\WeZu} \refmark{\rMan,\rWeZu}
$ x^i x^j = q^{-1} {R^{ji}}_{kl} x^k x^l $ and
$ y_i y_j = q^{-1} y_k y_l {R^{kl}}_{ji} $
invariant w.r.t. the QG coaction
$ {x'}^i = {T^i}_j x^j $ and $ {y'}_i = y_j {(T^{-1})^j}_i $
respectively. Commutation relations between $x^i$ and $y_j$ can be
determined using \rtt\ as $ x^i y_j = ({\it const.})\  y_k x^l
{R^{ik}}_{lj} $ and hence those of their product
$ {K^i}_j = x^i y_j $ which coincide with \alg. This also gives a
better understanding of the comodule property $K_T = T K T^{-1}$
of the RE.  \nextline
{\it (v)\phantom{iii}} If we impose suitable reality conditions on
$x^i , y_j $ and hence $ {K^i}_j $ then a linear combination of the
elements of \alg\  is just the $q$-deformed Minkowski space
\REF\rCWSSW{\CWSSW} \REF\rSmWZ{\SmWZ} \refmark{\rCWSSW,\rSmWZ}, where
$c_1$ is the time coordinate and $c_2$ the invariant length. Various
reality conditions are discussed in \refmark{\rSklb}, they parallel
those of \sl2. \nextline
{\it (vi)\phantom{ii}} Truncation of algebra \alg\  by $c_1 = 0$ can
be shown to lead to the quantum 2-sphere, a quantum analogue
of homogeneous spaces \Ref\rPod{\Pod}. \nextline
{\it (vii)\phantom{i}}  It is possible to introduce an index free
notation for quantum planes and extend it to the $K$-matrix, such that
the RE can be rewritten in exchange algebra form with four $R$-matrices
on one side \REF\rMada{\Mada} \refmark{\rSklb,\rMada}. \nextline
{\it (viii)} The monodromy $M = P exp\bigl({2\pi i \over k}
\int^{2\pi}_0 J(x) \, dx \bigr)$ of the \sl2 Kac-Moody current
satisfies the RE when regularized on a one-dimensional lattice with
periodic boundary conditions \Ref\rAFS{\AFS}. As this commutation
relation of the regularized monodromy holds for arbitrary numbers of
sites it might be expected to survive the continuum limit.
\nextline
{\it (ix)\phantom{ii}} Neither $K^{-1}$ nor $K^2$ is a solution of the
RE.

The last remark leads us to a very important property of the RE,
namely, given two {\it different} solutions of the RE satisfying a
certain compatibility condition then one can use them to construct new
solutions \refmark{\rKulb,\rKuSa,\rSch}. Explicitly, let $K$ and $K'$ be
solutions of \re\ then
$$ (i) \quad {\widetilde K} = K K'   \qquad  {\rm and} \qquad
(ii) \quad {\widetilde {\widetilde K}} = K K' K^{-1} \eqn \prod $$
are also solutions of \re\  provided $K$ and $K'$ commute as follows
$$  R K^{\phantom *}_1 R^{-1}K'_2 = K'_2 R K^{\phantom *}_1 R^{-1} .
\eqn \retwi $$
This equation is invariant under the coaction $ K_T = T K T^{-1} $
and $ K'_S = S K' S^{-1} $ if $S$ also obeys QG relations \rtt\ and in
addition $ R T_1 S_2 = S_2 T_1 R $, especially we can put $S$
equal to $T$.
Note that the second composite solution in \prod\ is not a trivial
consequence of the first since $\Ri$ does not solve the RE. This
process of building up new solutions can obviously be continued using
newly constructed solutions if they satisfy \retwi, but some care has
to be taken to keep track of the ordering as \retwi\ is not symmetric
under exchange of $K$ and $K'$. This will become clearer when we
discuss systems of solutions of both \re\ and \retwi. We will
sometimes refer to both equations as extended RE algebra.
Equation \retwi\ gives 16 commutation relations between the elements
of $K$ and $K'$
$$ \eqalign{a' a &= a a' - q \w b c' , \cr
            a' b &= b a' , \cr
            a' c &= c a' + q \w (a - d) c' , \cr
            a' d &= d a' + q^{-1} \w b c' , \cr
            b' a &= a b' + q \w b (a' - d') , \cr
            b' b &= q^2 b b' , \cr
            b' c &= q^{-2} c b' + (1 + q^{-2}) \w^2 b c' \cr
                 &{\phantom =}\  - q^{-1} \w (a - d) (a' - d') , \cr
            b' d &= d b' - q^{-1} \w b (a' - d') , \cr}  \qquad
\eqalign{c' a &= a c' , \cr
         c' b &= q^{-2} b c' , \cr
         c' c &= q^2 c c' , \cr
         c' d &= d c' , \cr
         d' a &= a d' + q^{-1} \w b c' , \cr
         d' b &= b d' , \cr
         d' c &= c d' - q^{-1} \w (a - d) c' , \cr
         d' d &= d d' - q^{-3} \w b c', \cr} \eqn \recom $$
and they only depend on $q^2$. Note that $K$ and $K'$ are commuting
for $q=1$ even if one linearizes them, whereas the RE in this
case produces the undeformed $sl(2)$ Lie algebra relations
\REF\rJur{\Jur} \REF\rZum{\Zum} \refmark{\rJur,\rZum}.
The extended \rea\ was implicitly contained also in
the construction of complex quantum groups recently
\Ref\rCDSWZ{\CDSWZ}, where relations \recom\ describe commutation
relations among the generators of the quantum algebra and their
complex conjugates, while both sets individually satisfy \alg.
Algebra \alg\ was also constructed in \refmark{\rMada} in the
framework of braided tensor categories.\footnote\dag{This approach to
the RE algebra takes the point of view that one has an extra braiding
between elements of the two copies of the algebra in the coproduct
$\Delta (K) = K {\dot \otimes} K$ s.t. $ (1 \otimes a) \cdot (a
\otimes 1) = a \otimes a - q \w b \otimes c $, for example, note the
new term on the RHS.  Denoting $1 \otimes a$ as $a'$ and $a \otimes 1$
as $a$, etc., this relation is identical with the first of \recom. Then
this `braided coproduct' for the \rea\ is compatible with the algebra
relations and is of the same form as for the QG and quantum algebra.
This so-called `braided group' which seemingly can be associated to
any QG was cast into RE form in \Ref\rMadb{\Madb}.} \nextline
An important point is further that the central elements of $K$ and $K'$
are mutually central in both algebras, i.e.
$$ [{K^i}_j,c'_m] = [{{K'}^i}_j,c_m] = 0 , \quad m=1,2 . \eqn \mutcen $$
It is obvious that we have central elements for the combined solutions
and also characteristic equations, for example
$$ q K K' + q^{-1} (K K')^{-1} - C_1 I = 0 , \eqn \comcen $$
where $ C_1 = q^{-1} (a a' + b c') + q (c b' + d d') $.

It is known that the RE algebra has a representation in terms of the
quantum algebra generators \Ref\rMoRe{\MoRe}. The \sl2 algebra
dual to the QG \rtt\  can similarly be written in  matrix form
\refmark{\rFRT}
$$ \Rt L^{\e_1}_1 L^{\e_2}_2 = L^{\e_2}_2 L^{\e_1}_1 \Rt , \qquad
(\e_1,\e_2) \in  \{(+,+),(+,-),(-,-)\}  \eqn \rll  $$
where
$$ L^+ = \pmatrix{q^{H/2} & q^{-1/2} \w X^- \cr 0 & q^{-H/2} \cr} ,
\qquad L^- = \pmatrix{q^{-H/2} & 0 \cr -q^{1/2} \w X^+ & q^{H/2} \cr}
\eqn \lplm $$
and this gives the \sl2 algebra
$$ [H,X^{\pm}] = \pm 2 X^{\pm} , \qquad
   [X^+,X^-] = \w^{-1} (q^H - q^{-H})  \eqn \sltwo  $$
with antipode $ S(H) = - H $, $ S(X^{\pm}) = - q^{\mp 1} X^{\pm} $
and coproduct $ \Delta(L^{\pm}) = L^{\pm} {\dot \otimes} L^{\pm} $.
It is easy to show using \rll\  that $ K = S(L^-) L^+ $ represents a
solution of the RE, explicitly $K$ is given by
$$ K = \pmatrix{q^H & q^{-1/2} \w q^{H/2} X^- \cr
q^{-1/2} \w X^+ q^{H/2} & q^{-H} + q^{-1} \w^2 X^+ X^- \cr} .
\eqn \lsol $$
This representation of the RE algebra has quantum determinant $c_2=1$
and the quantum trace $c_1$ is just the quadratic Casimir operator of
the quantum algebra \sl2.
 
We now find whole towers of new representations of the RE algebra in
terms of quantum algebra generators and generalize them to the
extended RE algebra. They will be useful for representing the braid
group \bralgf. To avoid clumsy notation we introduce the abbreviation
$S^{\pm} \equiv S(L^{\pm})$.
There is a simple way to produce further representations of the RE
algebra, namely by means of the coproduct $\D$ which gives
representations on tensor products of spaces, in the simplest case of
two spaces we obtain
$$ \eqalign{\D({K^i}_j) &= \D({{S^-}^i}_k) \D({{L^+}^k}_j)  \cr
   &= {K^m}_n \otimes {{S^-}^i}_m {{L^+}^n}_j  \cr
   &= {(1 {\dot \otimes} S^-)^i}_m {(K {\dot \otimes} 1)^m}_n
      {(1 {\dot \otimes} L^+)^n}_j  \cr
   &\equiv {{S^-_2}^i}_m {{K^{\phantom -}_1}^m}_n {{L^+_2}^n}_j}
\eqn \copr $$
or in matrix notation simply $\D(K) = S^-_2 K^{\phantom -}_1 L^+_2$.
Note that this coproduct for $K$ is a consequence of the one for
$L^{\pm}$ and cannot be expressed as $\D(K) = K_1 K_2$ (cf. footnote
after eqn. \recom). We stress that whenever we write down tensor
products in the following then entries of different spaces are
strictly commuting. \nextline
We can get a whole string of solutions of the \rea\
generalizing \copr\ by repeatedly applying the coproduct,
and in addition embed it into the $g$-fold tensorproduct of the
universal enveloping algebra of \sl2 leading to the definition
$$ \k0m = \h-g \cdots \h-{g-(m-2)} K^{}_{g-(m-1)}
\l+{g-(m-2)} \cdots \l+g ,  \qquad 1 \leq m \leq g  \eqn \kzerm $$
where $K^{\phantom -}_i = \h-i \l+i$ and
$ \l+i = 1 {\dot \otimes} \cdots 1 {\dot \otimes}
L^+ {\dot \otimes} 1 \cdots {\dot \otimes} 1 $ with $L^+$ inserted
into the $i$-th position, etc. In the simplest case of \sl2 all
objects on the RHS of \kzerm\ are $2 \times 2$ matrices, matrix
multiplication being understood, and their entries take values in
the  $g$-fold tensor product $ U_q^{\otimes g} (sl(2)) $. So \kzerm\
defines $g$ operators $\k01 = K^{\phantom -}_g,\  \k02 = \h-g
K^{\phantom -}_{g-1} \l+g,\  \ldots,\  \k0g = \h-g \cdots \h-2
K^{\phantom -}_1 \l+2 \cdots \l+g$. We keep
$g$ arbitrary but fixed, it will correspond to the genus of the
handlebody later on. Formally one may envisage the limit $g
\rightarrow \infty$ as one has a natural sequence of embeddings of
tensor products into higher ones. Incidentally, we found that a
variant of the operators $\k0m$ has been employed in the formulation
of quantum differential geometry \Ref\rSWZ{\SWZ}.
\nextline
However, there is a drawback because $\k0m$ and $\k0n$ do
not satisfy \retwi\ for $m \not= n$ but instead again the RE
$$ R (\k0m)_1 \Rt (\k0n)_2 = (\k0n)_2 R (\k0m)_1 \Rt , \qquad
m \geq n  \eqn \regen $$
and hence $\k0m$ cannot be used to represent the generators $\t_\a$.
Fortunately, this construction gives us a hint how to solve the
problem. We define two more sets of operators $\k+m$ and $\k-m$ by
$$ \k{\pm}m = \h{\pm}g \cdots \h{\pm}{g-(m-2)} K^{\phantom -}_{g-(m-1)}
\l{\pm}{g-(m-2)} \cdots \l{\pm}g , \eqn \kpmm $$
they are also solutions of the RE and have the following commutation
relations
$$ \eqalign{R (\k{\pm}m)_1 \Rt (\k{\pm}m)_2 &= (\k{\pm}m)_2 R
(\k{\pm}m)_1 \Rt , \cr
R (\k+m)_1 \Ri (\k+n)_2 &= (\k+n)_2 R (\k+m)_1 \Ri , \qquad m < n \cr
R (\k-m)_1 \Ri (\k-n)_2 &= (\k-n)_2 R (\k-m)_1 \Ri , \qquad m > n \cr
R (\k-m)_1 \Ri (\k+n)_2 &= (\k+n)_2 R (\k-m)_1 \Ri , \qquad m \not= n}
\eqn \recomgen $$
corresponding to \re\ and \retwi\ while the last equation gives the
commutation relations between the two sets of operators. They also
have definite commutation relations with $\k0m$ given by
$$ \eqalign{R (\k0m)_1 \Rt (\k{\pm}n)_2 &= (\k{\pm}n)_2 R
(\k0m)_1 \Rt , \cr
R (\k0m)_1 \Ri (\k+n)_2 &= (\k+n)_2 R (\k0m)_1 \Ri ,  \cr
R (\k-m)_1 \Ri (\k0n)_2 &= (\k0n)_2 R (\k-m)_1 \Ri .  \cr}
\qquad \eqalign{&m \geq n \cr &m < n \cr &m > n \cr}
\eqn \regentwi $$
These relations can be verified because we obviously have
$$ \eqalign{\Rt (\l{\e_1}m)_1 (\l{\e_2}m)_2 &= (\l{\e_2}m)_2
(\l{\e_1}m)_1 \Rt , \cr
(\l{\e_1}m)_1 (\l{\e_2}n)_2 &= (\l{\e_2}n)_2 (\l{\e_1}m)_1 ,
\qquad \quad m \not= n \cr} \eqn \rllgen $$
as a consequence of \rll. Thus there are two sets of operators
expressed in terms of quantum algebra generators that can be used to
represent the braid group generators $\t_\a$. Their meaning will be
clarified in the next section. In principle \recomgen\ constitutes
an infinite dimensional algebra and it might have representations
other than by quantum algebra generators.
Some of the relations in \recomgen\ and \regentwi\ have the form of
\retwi\ and so the new solutions of the RE in \prod\ can be built.
For example, if one considers only the $\k+m$ series then it can be
seen that the product $\k+m \k+n$, $n>m$, has also commutation
relation \retwi\ with $\k+p$ if $p>n$, and this behaviour persists for
operators with appropriate ordering. This is most easily seen in terms
of diagrams to be introduced in the next section. \nextline
For these new solutions (but not for $\k0m$) the characteristic
equation \cheq\ holds as well
$$ q \k{\pm}m + q^{-1} (\k{\pm}m)^{-1} - c_1 I = 0 ,
\eqn \chargen $$
with $ (\k{\pm}m)^{-1} = \h{\pm}g \cdots \h{\pm}{g-(m-2)}
K^{-1}_{g-(m-1)} \l{\pm}{g-(m-2)} \cdots \l{\pm}g $ and $K^{-1}_i =
\h+i \l-i$.  The central element $c_1$ as the quadratic casimir
operator of \sl2 remains unchanged. Finally, as representations of the
extended \rea\ their commutation relations are automatically invariant
w.r.t. the comodule transformation $({\k{\pm}m})_T = T \k{\pm}m
T^{-1}$.

We remark that instead of \copr\ we could have used the permuted
coproduct $\D'(K) = \h-1 K^{\phantom{+}}_2 \l+1$ having a
generalization as $\k0m = \h-1 \cdots \h-{m-1} K^{\phantom{-}}_m
\l+{m-1} \cdots \l+1$, and hence $\k{\pm}m = \h{\pm}1 \cdots
\h{\pm}{m-1} K^{\phantom{\pm}}_m \l{\pm}{m-1} \cdots
\l{\pm}1$. They correspond to a reverse ordering of spaces and
differ from \kpmm, but do also satisfy \regen,\ \recomgen\ and
\regentwi. However, $\k{\pm}m$ in this simpler form is not consistent
with our conventions in the next section, so we do not consider this
further.

The RE algebra therefore plays different roles, it is a comodule w.r.t.
the QG and on the other hand it acts via \lsol\ on representations of
the quantum algebra dual to the QG.  Further applicatons of the \rea\
were mentioned in \refmark{\rKSS}.


\chapter{REPRESENTATIONS OF THE BRAID GROUP ON HANDLEBODIES}

The braid group $B^g_n$ on a solid handlebody $H_g$ of genus $g$ was
described in \refmark{\rSos}. In addition to the generators $ \s_i ,
i=1,\ldots,n-1 $ of the braid group $B_n$ defined on a 3-dimensional
manifold of genus zero there are generators $ \t_\a , \a=1,\ldots,g $
implementing windings around the $g$ handles. The  algebra
is given by
$$\eqalign{\s_i \s_{i+1} \s_i &= \s_{i+1} \s_i \s_{i+1} , \quad
           i=1,\ldots,n-1 \cr
\s_i \s_j &= \s_j \s_i , \quad \mid i-j \mid \geq 2 \cr
\s_i \t_\a &= \t_\a \s_i , \quad i \geq 2 , \ \a =1,\ldots,g  \cr
\s_1 \t_\a \s_1 \t_\a &= \t_\a \s_1 \t_\a \s_1 , \cr
\s_1 \t_\a \s_1^{-1} \t_\b &= \t_\b \s_1 \t_\a \s_1^{-1} ,
\quad \a < \b  \cr} \eqn \bralg $$
and the first two relations define the well known Artin braid group
\refmark{\rArt}. We refer to \refmark{\rSos} for  details and
references, here we only explain conventions briefly which should
make \bralg\ fairly transparent.

On the handlebody (Fig.1) it is possible, without loss of generality,
to prescribe a fixed ordering of the points where the strands begin
(resp. end) having coordinates $ P_i^{(1)} = ({i \over {n+1}},{1 \over
2},1)$ \ (resp. $ P_j^{(0)} = ({j \over {n+1}},{1 \over 2},0)$,
\ $i,j=1,\ldots,n$ in a lefthanded $(x,y,z)$-coordinate system. So the
unit cube in the positive octant is contained in $H_g$ and the usual
braids are obtained by connecting points $P_i^{(1)}$ and $P_j^{(0)}$
by strands confined to the unit cube. The braid diagram is obtained by
projecting on the $x$-$z$-plane.  The handles are positioned, say, to
the left of the unit cube around coordinates $ h_\a = ({-\a \over
{g+1}},y,1-{\a \over {g+1}}), \ \a =1,\ldots,g $. For the braid group
on $H_g$ the strands are allowed to leave the unit cube at height $z$
and go around the handle $h_\a $ counterclockwise for $\t_\a $
(clockwise for $\t_{\a}^{-1}$) and then come back to the unit cube at
height $z-\d$, $\d$ small. The convention\footnote\dag{We have chosen
conventions slightly different from \refmark{\rSos}, especially in the
last equation of \bralg\ the condition in \refmark{\rSos} is $\a > \b$,
and also `strands should be over if $z_2 > z_1$'.} is that strands
leaving or entering the unit cube at height $z_1$ should be over those
doing so at $z_2$ in the projection onto the $x$-$z$-plane if $z_1 >
z_2$.  Within the unit cube strands can only go downward in the
negative $z$-direction. This definition can be further formalized, but
everything is rather intuitive.
\vskip1cm \vbox{
\centerline{\psfig{figure=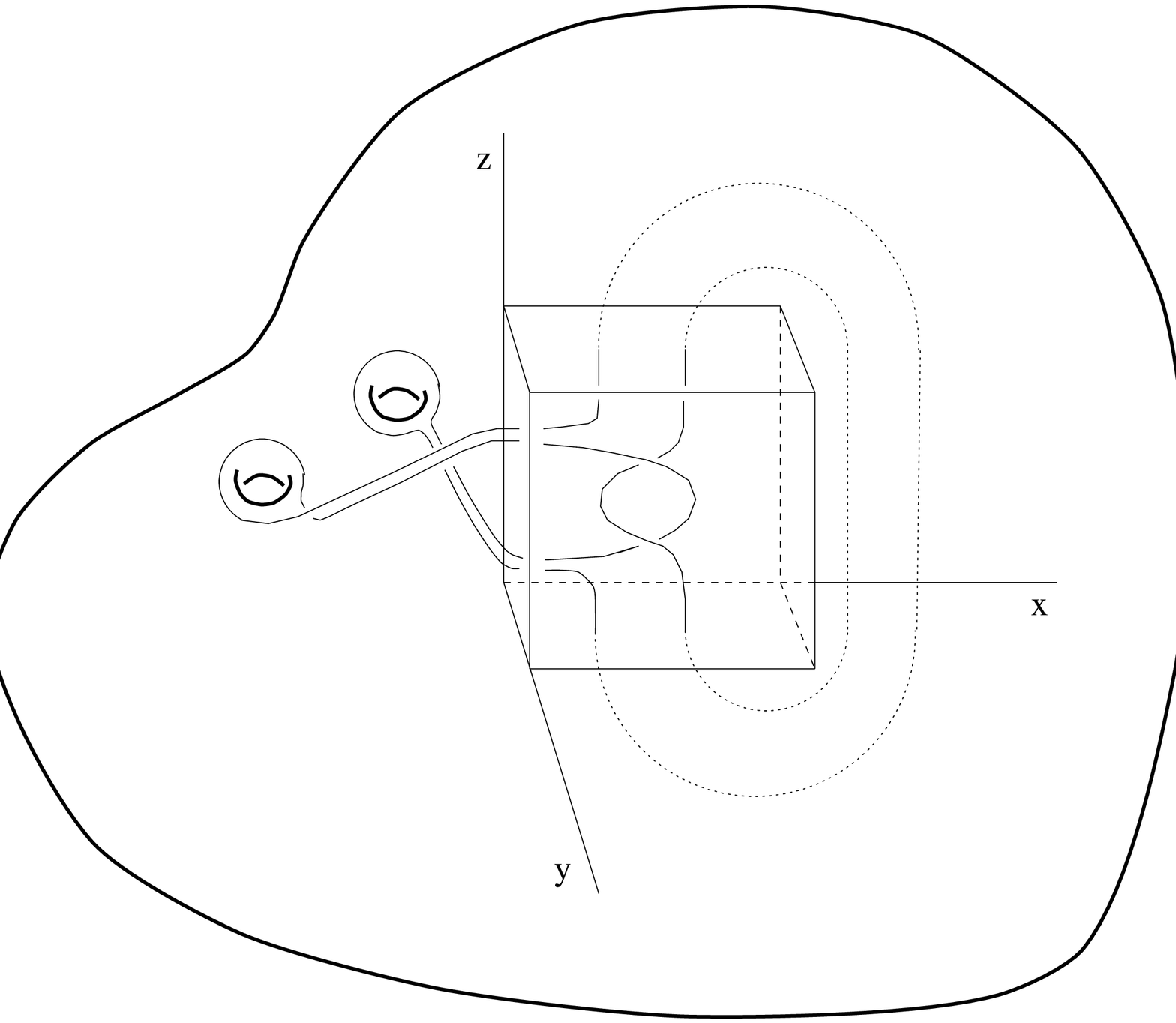,width=10cm}}
\centerline{Fig.1: \  The 2-braid $\t_2^{-1} \s_1^2 \t_1$ and its
closure (dotted lines)}}
\vskip1cm  \noindent
For our arguments it is more appropriate to think of piercing long
bars through the handles and after that forget about them. Then, if
we rotate the bars by $\pi / 4$ around the $x$-axis
counterclockwise to $h'_\a = ({-\a \over {g+1}},{\a \over {g+1}}-1,z)$
we can depict the braiding in a more systematic way by projecting on
the $x$-$z$-plane. For example, the fourth equation of \bralg\ can be
represented graphically as in Fig.2 where, as usual, $\s_1$ has been
represented by a crossing of two strands.
\vskip1cm \vbox{
\centerline{\psfig{figure=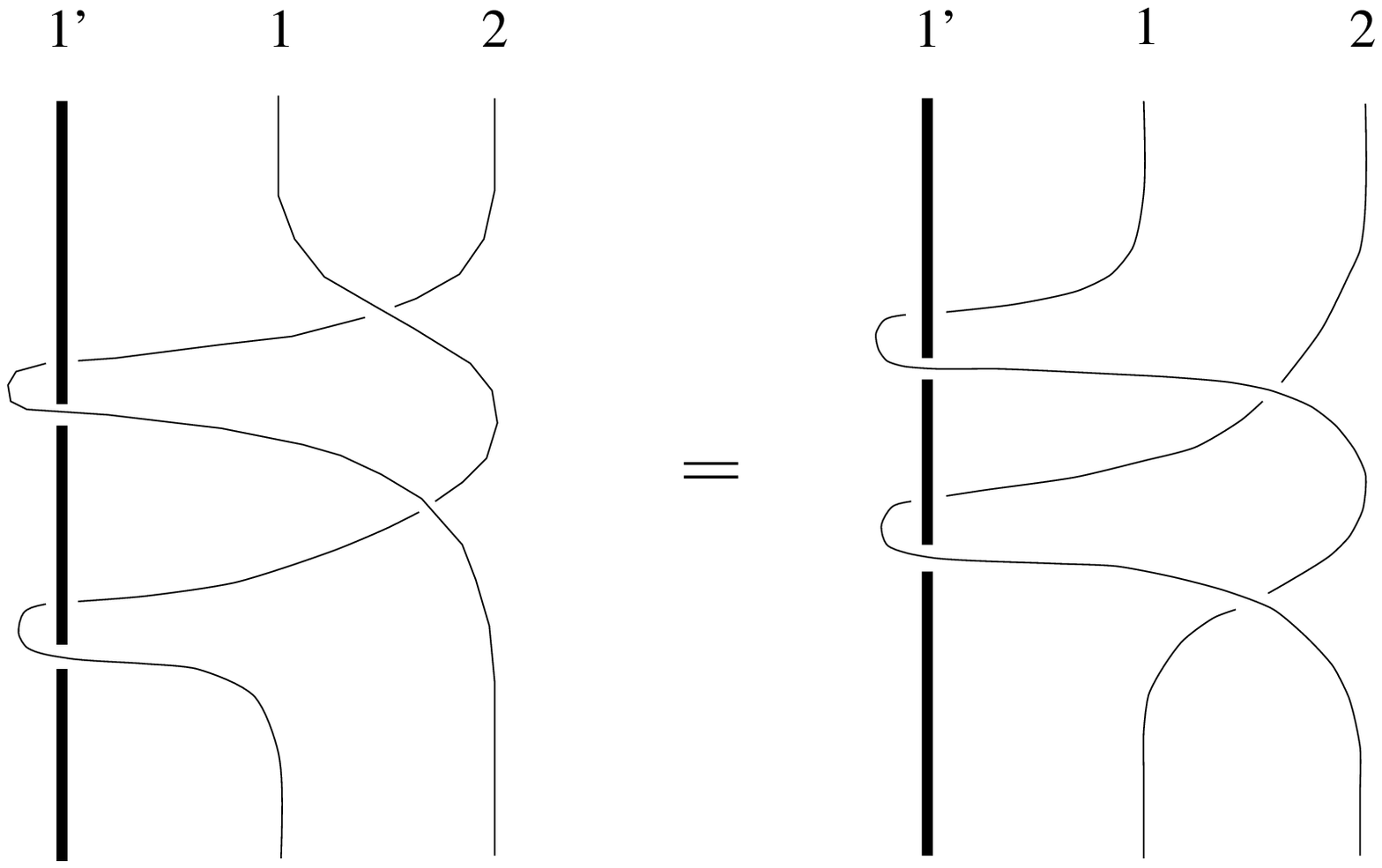,width=8cm}}
\centerline{Fig.2: \  Graphical representation of the
reflection equation (for later}
\centerline{convenience numbering of spaces is indicated)
{\phantom {xxxxx}}}}
\vskip0.5cm \noindent
Similarly, the last equation of \bralg\  can be represented as in
Fig.3 and is proven easily this way by pulling lines appropriately.
\vskip1cm \vbox{
\centerline{\psfig{figure=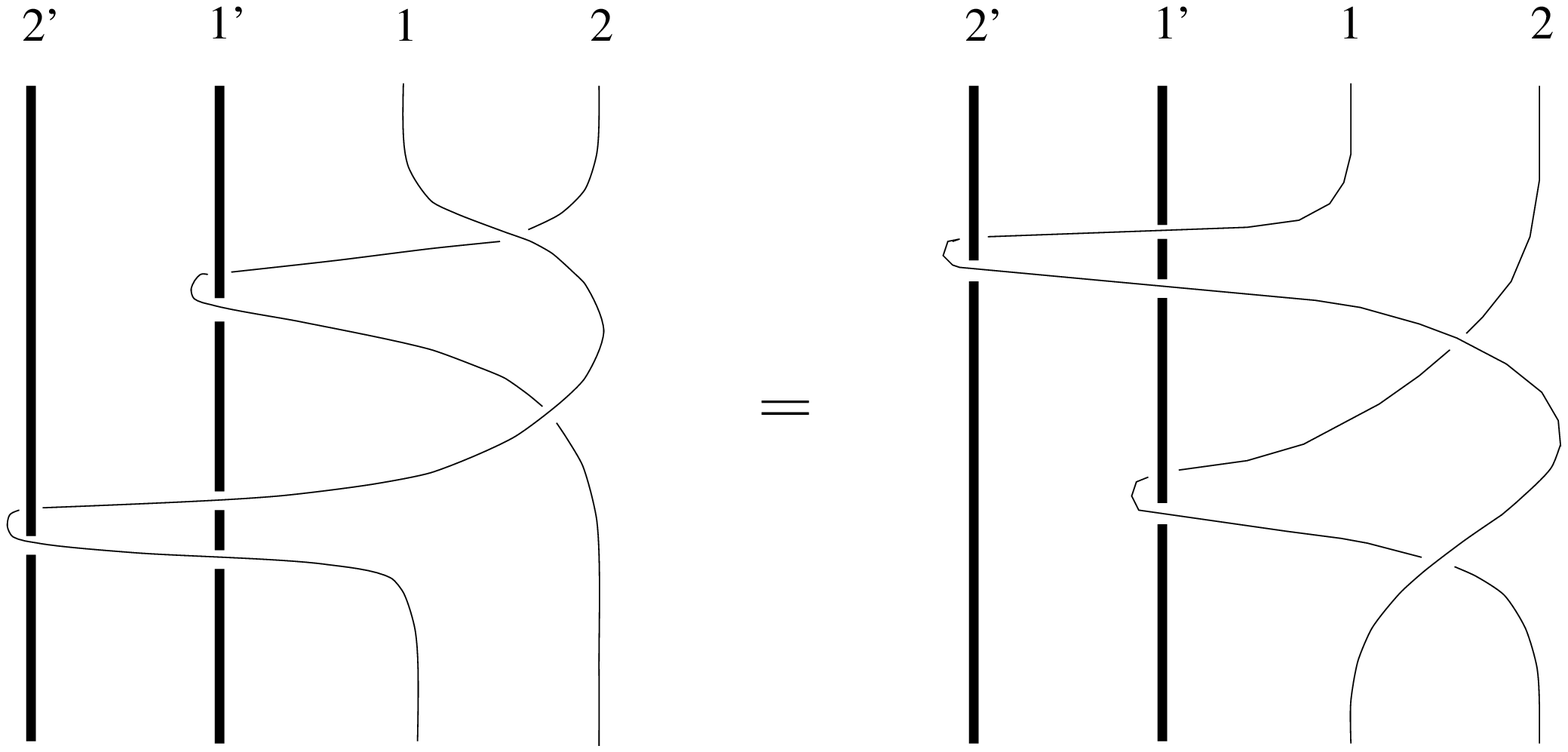,width=10cm}}
\centerline{Fig.3: \  Graphical representation of compatibility
condition \retwi}}
\vskip0.5cm
The strands leaving the unit cube to wind around the bars always
belong to the first space $V_1$ of the tensor product $ V(n) = V_1
\otimes \cdots \otimes V_n$ on which the $\s_i$ act, and this explains
why only $\s_1$ is non-commuting with $\t_\a $. It also means that
$\t_\a$ is acting non-trivially only in $V_1$ by some operator
$K_{(\a)}$, so we put
$$ \eqalign{\s_i &= q\  1 \otimes \ldots 1 \otimes \Rh_{i,i+1} \otimes
1 \ldots \otimes 1, \quad i=1,\ldots,n-1 \cr
\t_\a &= q^3\   K_{(\a)} \otimes 1 \ldots \otimes 1 , \quad
\a=1,\ldots,g \cr} \eqn \ident $$
where as usual $\s_i$ is acting non-trivially only in $V_i \otimes
V_{i+1}$ as $\s_i = P R_{i,i+1} \equiv \Rh_{i,i+1}$. The factor $q^3$
is inserted for later convenience to match the factor $q$ in the
definition of $\s_i$. Using this we can show explicitly that \bralg\
is equivalent to the Yang-Baxter equation and to the extented \rea,
by identifying $\s_1 = \Rh_{12}$ and $\t_\a = (K_{(\a)})_1$, plus
two other rather obvious consistency conditions as given in \bralg.
This equivalence is also explained in \refmark{\rKulb,\rKuSa}.

Thus $\s_i$ has an explicit matrix representation, but what about
$\t_\a$? Because $\s_i$ is represented by a $R$-matrix one should
expect that the same holds true for $\t_\a$, and this is supported
by Fig.2 which suggests to represent the effect of a handle on a
strand going around it by the square of a $R$-matrix. Let us consider
first the genus one case, i.e. only the RE has to be taken into
account. As shown in the previous section $ K = S^- L^+ $ is a
solution of the RE and this tells us how to represent $\t_\a$ because
$S^-$ and $L^+$ are related to the $R$-matrix. In fact, looking at
the universal $R$-matrix of \sl2 acting on $V_1 \otimes V_2$
$$ R_U = q^{{1 \over 2} H \otimes H} \sum^\infty_{n=0} {(1-q^{-2})^n
\over \lbrack n;q^{-2} \rbrack !}\bigl(q^{-{1 \over 2}H} X^- \bigr)^n
\otimes \bigl(q^{{1 \over 2} H} X^+ \bigr)^n , \qquad
\lbrack n;q \rbrack = {(1-q^n) \over (1-q)}  \eqn \ru  $$
we can represent the \sl2 generators either on $V_1$ or $V_2$
(the fundamental representation of \sl2 is
{\hbox{$\rho_f(H)=$  $1\, \ 0\ \choose 0\,-1$}},
{\hbox{$\rho_f(X^+)=$  $0\,1\choose 0\,0$}},
{\hbox{$\rho_f(X^-)=$  $0\,0\choose 1\,0$}}) giving
$$ \eqalign{\rho_f(R_U)\big\vert_{V_1} &=
\pmatrix{q^{H/2} & 0 \cr q^{-1/2} \w X^+ & q^{-H/2} \cr} = S^- , \cr
\rho_f(R_U)\big\vert_{V_2} &=
\pmatrix{q^{H/2} & q^{-1/2} \w X^- \cr 0 & q^{-H/2} \cr} = L^+ .}
\eqn \alrep  $$
And further representing in a second step the `semiuniversal'
operators $S^-$ and $L^+$ we get
$$ \rho_f(S^-) = q^{-1/2} R , \qquad \rho_f(L^+) = q^{-1/2} \Rt , \qquad
\rho_f(S^- L^+) = q^{-1} {\Rth}{}^2  \eqn \oprep $$
where $\Rth = P \Rh P = R P$. This means that we have to represent a
strand `interacting' with the handle like in Fig.2 by
$ {K_{(1)}{}^i}_j = q^{-1} ({\Rth}{}^2{}^i{}_j)^m{}_n \equiv q^{-1}
({\Rh}^2{}^m{}_n)^i{}_j$, where the $(ij)$ indices are in the first
space of $V(n)$ and therefore $ {K_{(1)}{}^i}_j = q^{-1}
({\Rh}^2{}^m{}_n)^i{}_j$  suits the graphical representation in Fig.2.
In effect we have translated a topological property of the solid torus
into a quantum algebra operator acting on $V_1$ and an additional
`internal' space $V_{1'}$ embedded into the space $V(1;n) = V_{1'}
\otimes V_1 \otimes \cdots V_n$. As a consequence we have a
two-dimensional representation of the \rea\  given by matrices
$$ \eqalign{{a^m}_n &= \pmatrix{q & 0 \cr 0 & q^{-1} \cr} , \cr
{c^m}_n &= \pmatrix{0 & q^{-1} \w \cr 0 & 0 \cr} ,} \qquad  \quad
\eqalign{{b^m}_n &= \pmatrix{0 & 0 \cr q^{-1} \w & 0 \cr} , \cr
{d^m}_n &= \pmatrix{q^{-1}(1 + \w^2) & 0 \cr 0 & q \cr} ,}
\eqn \twdrep $$ \break
which indeed satisfy \alg. The operator $ \k{}1 = S^- L^+
\vert_\rho $ appeared also in the context of conformal field theory
\refmark{\rMoRe} and was used there, for example, in connection with
topology changing amplitudes in Chern-Simons field theory.

\noindent It is easy to check from \rmat\ that both $\Rh$ and $\Rth$
obey a quadratic equation of the form
$$ \Rh^2 - \w \Rh - I = 0 . \eqn \rchar $$
This, in turn, leads to a quadratic equation for $\Rh^2$
$$ \Rh^2 + \Rh^{-2} - (q^2 + q^{-2}) I = 0 , \eqn \rsqch $$
and this is nothing but the characteristic equation \cheq\ for
$K_{(1)} = q^{-1} \Rh^2_{1' 1}$. Comparing with \cheq\ we can identify
$c_1(K) = q^2 + q^{-2}$ (discarding the $2 \times 2$ identity matrix),
and this can be verified by calculating the quantum trace of the
explicit representation \twdrep. Hence, for this representation of
$K_{(1)}$ its characteristic equation follows from the one for the
$R$-matrix. Therefore the somewhat ad hoc assumption of considering
the characteristic equation of the $K$-matrix as a skein relation for
lines going around handles in \refmark{\rSch} is justified because
\rchar\ has an interpretation as a skein relation. Even more so as we
have seen that handles themselves can be considered as kind of lines
in topologically trivial regions, we will focus on this in the next
section.

In order to explain the case of arbitrary genus it is sufficient to
look at $g=2$. Now we also need to take into account the last equation
of \bralg. From Fig.3 we can guess that $K_{(1)}$ is as before, but
$K_{(2)}$ should be represented by a product of four $R$-matrices
acting on $V(2;n) = V_{2'} \otimes V_{1'} \otimes V_1 \otimes
\cdots V_n$ as
$$ K_{(1)} = q^{-1} I^{\phantom{1}}_{2'} \otimes \Rh^2_{1'1} ,
\qquad  K_{(2)} = q^{-1} \Rh^{-1}_{1'1} \Rh^2_{2'1'}
\Rh^{\phantom{-1}}_{1'1} . \eqn \kontw  $$
The indices characterizing operators $a,b,c,d$ belong to $V_1$, all
primed indices refer to `internal' spaces  related to the handles of
the manifold. Thus we can read off from \kontw\ the explicit
four-dimensional representation using \rmat
$$  \eqalign{
a_{(1)} &= \pmatrix{q & 0 & 0 & 0 \cr 0 & q^{-1} & 0 & 0 \cr
                   0 & 0 & q & 0 \cr 0 & 0 & 0 & q^{-1} \cr}, \cr
c_{(1)} &= \pmatrix{0 & q^{-1} \w & 0 & 0 \cr 0 & 0 & 0 & 0 \cr
                   0 & 0 & 0 & q^{-1} \w \cr 0 & 0 & 0 & 0 \cr}, \cr
        &\phantom{=}  \cr
a_{(2)} &= \pmatrix{q & 0 & 0 & 0 \cr 0 & q & - \w^2 & 0 \cr
                   0 & 0 & q^{-1} & 0 \cr 0 & 0 & 0 & q^{-1} \cr}, \cr
c_{(2)} &= \pmatrix{0 & 0 & \w & 0 \cr 0 & 0 & 0 & q^{-2}\w \cr
                   0 & 0 & 0 & 0 \cr 0 & 0 & 0 & 0 \cr}, \cr}
\qquad \quad \eqalign{
b_{(1)} &= \pmatrix{0 & 0 & 0 & 0 \cr q^{-1} \w & 0 & 0 & 0 \cr
                   0 & 0 & 0 & 0 \cr 0 & 0 & q^{-1} \w & 0 \cr}, \cr
d_{(1)} &= \pmatrix{q^{-1}(1+\w^2) & 0 & 0 & 0 \cr 0 & q & 0 & 0 \cr
            0 & 0 & q^{-1}(1+\w^2) & 0 \cr 0 & 0 & 0 & q \cr}, \cr
        &\phantom{=}  \cr
b_{(2)} &= \pmatrix{0 & 0 & 0 & 0 \cr q^{-2}\w^2 & 0 & 0 & 0 \cr
            q^{-2}\w & 0 & 0 & 0 \cr 0 & \w & - \w^2 & 0 \cr}, \cr
d_{(2)} &= \pmatrix{q^{-1}(1+\w^2) & 0 & 0 & 0 \cr
                    0 & q^{-1}(1+\w^2) & q^{-2}\w^2 & 0 \cr
                    0 & 0 & q & 0 \cr 0 & 0 & 0 & q \cr}.
\cr} \eqn \fodrep  $$
These matrices do not only satisfy \alg\ but really
provide a non-trivial explicit representation of \recom\ (with
$K_{(1)} = K,\ K_{(2)} = K'$), proving that \kontw\ is indeed a
representation of \bralg. Furthermore, using the representation
$\rho_f$ of $H$ and $X^\pm$ it can be verified that \fodrep\ may
equally well be obtained from quantum algebra solution \kpmm\ of the
extended \rea
$$ K_{(1)} = \rho_f(K_2) \equiv \rho_f(\k+1) , \qquad
   K_{(2)} = \rho_f(\h+2 K^{\phantom{-}}_1 \l+2) \equiv \rho_f(\k+2) .
\eqn \kfotw $$
As shown in section 2 all quantum algebra solutions satisfy the same
characteristic equation, so $K_{(1)}$ and $K_{(2)}$ do satisfy
\rsqch\ which is obvious from \kontw\ anyway. Then one might wonder
what the meaning is of $\k-m$, it can be seen that it plays the
same role as $\k+m$ but corresponds to a different incompatible
set of conventions compared to those given in the beginning (i.e. $\a
> \b$ in \bralg, $z_2 > z_1$ for strands leaving or entering the unit
cube, clockwise rotation of bars corresponding to handles, interchange
of figures for $K$ and $K^{-1}$). This choice gives nothing new and
needs not to be considered, for example, $K_{(2)} = \rho_f(\k-2)$
is the same as in \fodrep\ but with all four matrices transposed and
$b_{(2)}$ interchanged with $c_{(2)}$. The last relation of \bralg\
in this case can be depicted as in Fig.4, the major difference being
some lines now going under the bars.
\vskip0.3cm \vbox{
\centerline{\psfig{figure=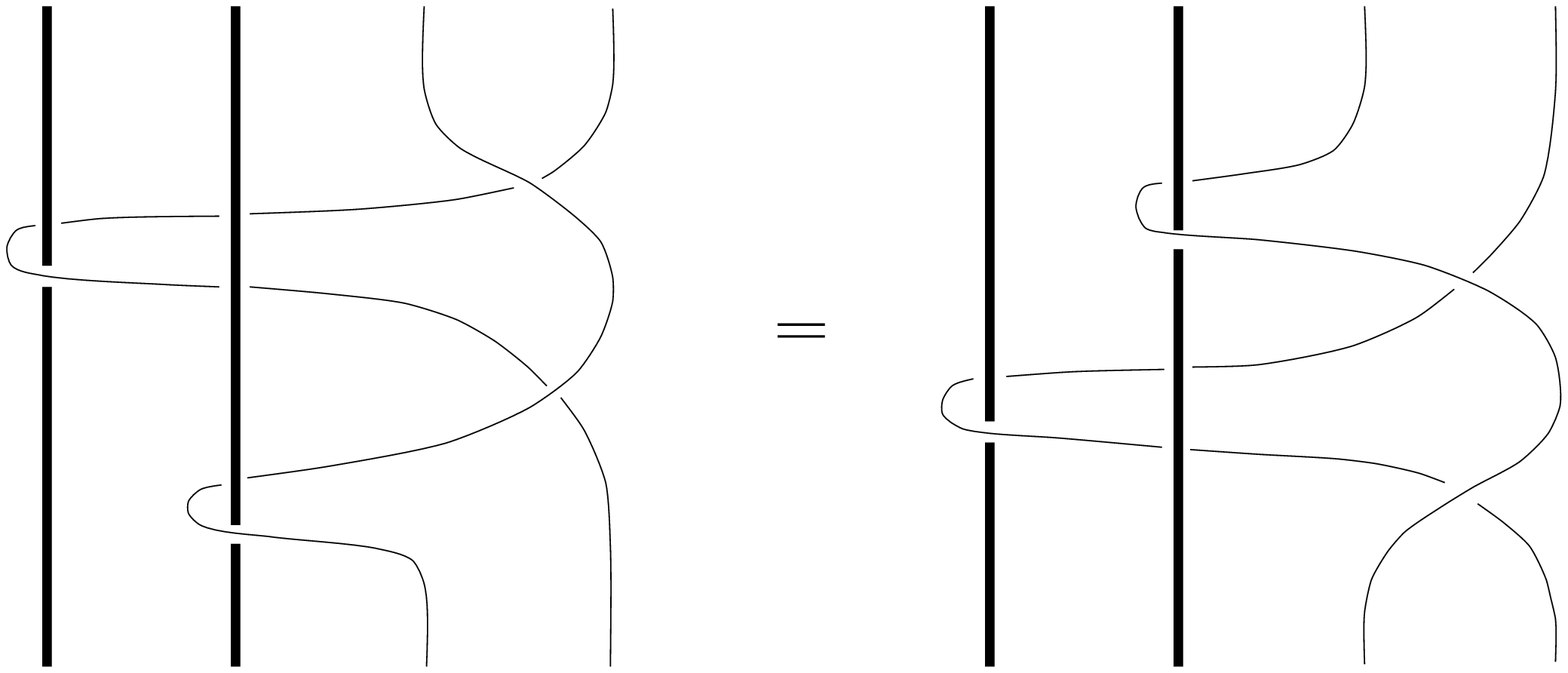,width=10cm}}
\centerline{Fig.4: \  Compatibility condition \retwi\ with $\t_\a$
represented by the $\k-m$ series}}
\vskip0.3cm
We still have to explain the meaning of \prod\ in terms of the braid
group generators. Property $(i)$ relates to successive application of
$K_{(1)}$ and $K_{(2)}$ leading to a new move encircling both bars as
displayed in Fig.5.
\vskip0.3cm \vbox{
\centerline{\psfig{figure=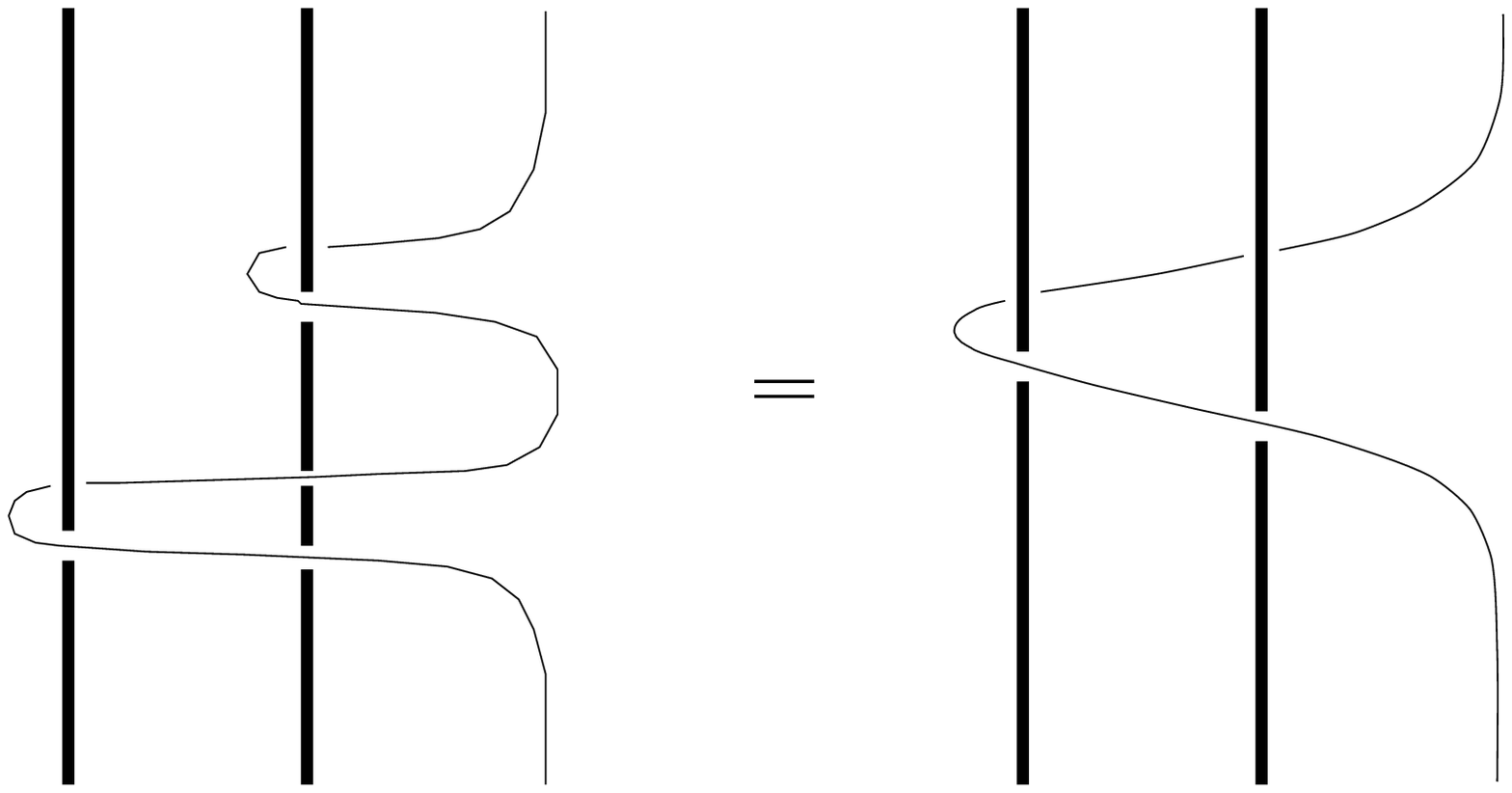,width=8cm}}
\centerline{Fig.5: \  The product $\t_1 \t_2$ (equivalent to
representing $\t_2$ by $\k02$)}}
\vskip0.3cm
\noindent
It is immediately clear that the move $K_{(1)} K_{(2)}$ again
satisfies the RE as can be seen just by inserting an additional bar
appropriately into Fig.2. Property $(ii)$ can be understood similarly
by looking at $\k{}1 \k{}2 \k{-1}1$ shown in Fig.6. If the results of
multiplying (products of) braid group generators obey the braid group
relations again it then graphically comes down to being able to `pull
lines' in a simple way.
\vskip0.3cm \vbox{
\centerline{\psfig{figure=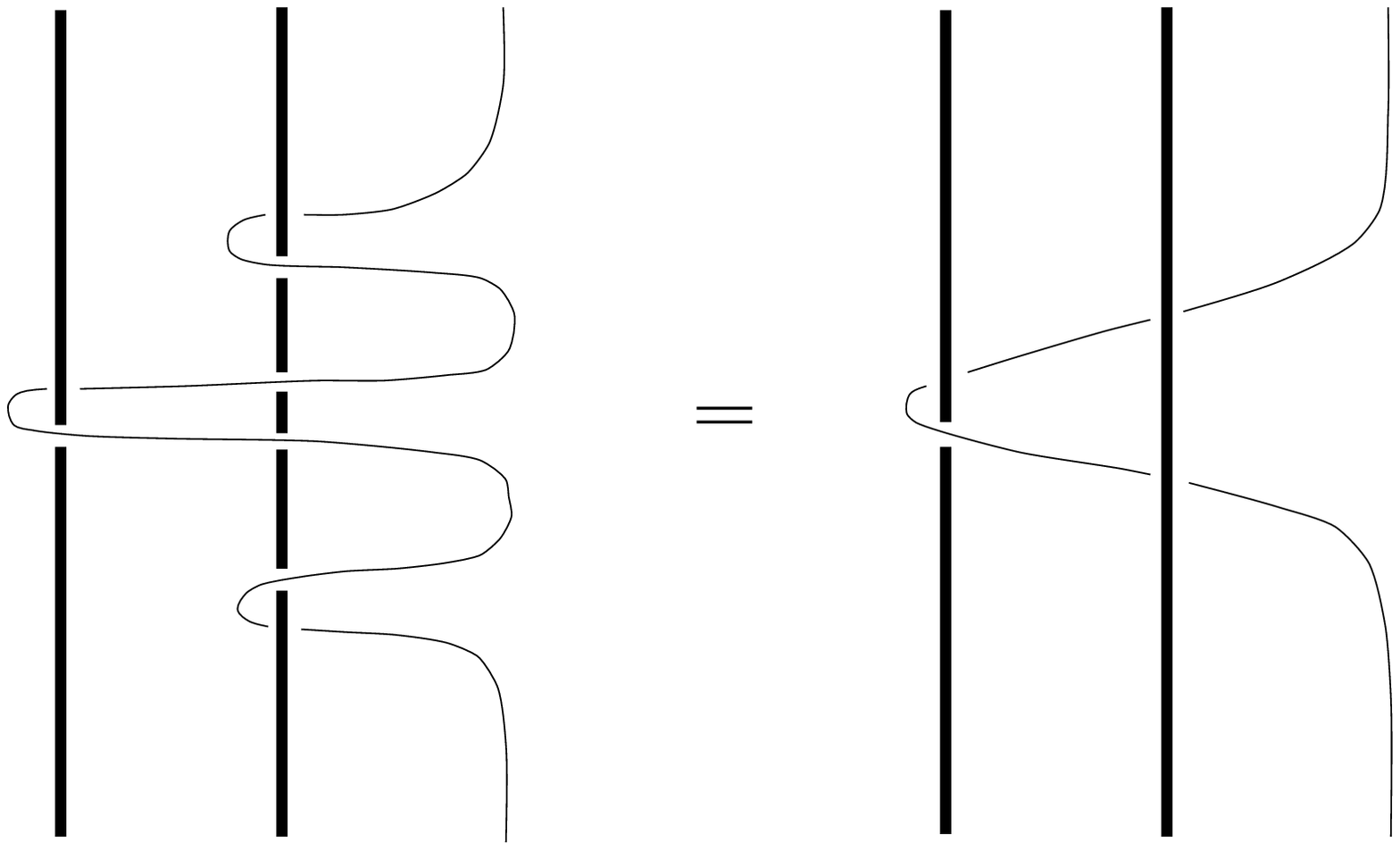,width=7cm}}
\centerline{Fig.6: \  The product $\t^{\phantom 1}_1 \t^{\phantom 1}_2
\t^{-1}_1$ (equivalent to representing $\t_2$ by $\k-2$)}}
\vskip0.5cm
\noindent
The results of figs.5,6 show that although we have chosen the $\k+m$
series to represent the braid group generators $\t_\a$, we now see the
$\k-m$ and $\k0m$ series appearing because the combined solutions are
precisely given by $\k02$ and $\k-2$ as can be checked by formulas.
Even though we start from $\k+m$ solely its properties as an extended
\rea\ force us to consider the whole system \regen,\ \recomgen\ and
\regentwi. As a side remark we mention here that if we may use graphs
for both $\k+m$ and $\k-n$ it is easy to see why they have no
commutation relation for $m=n$, it is just not possible to disentangle
the lines.

Now it should be clear how this generalizes to the case of arbitrary
genus. We will have $g$ bars corresponding to the handles and $\t_\a$
is represented by a strand going from first space over the first $(\a
- 1)$ bars to wind around the one corresponding to the handle $h_\a$
and going back again to $V_1$ over the first $(\a - 1)$ bars. An
example is shown in Fig.7. It is obvious that the $\t_\a$ satisfy
the defining relations \bralg\ and this can be proven analogously to
figs.3,4.
\vskip1cm \vbox{
\centerline{\psfig{figure=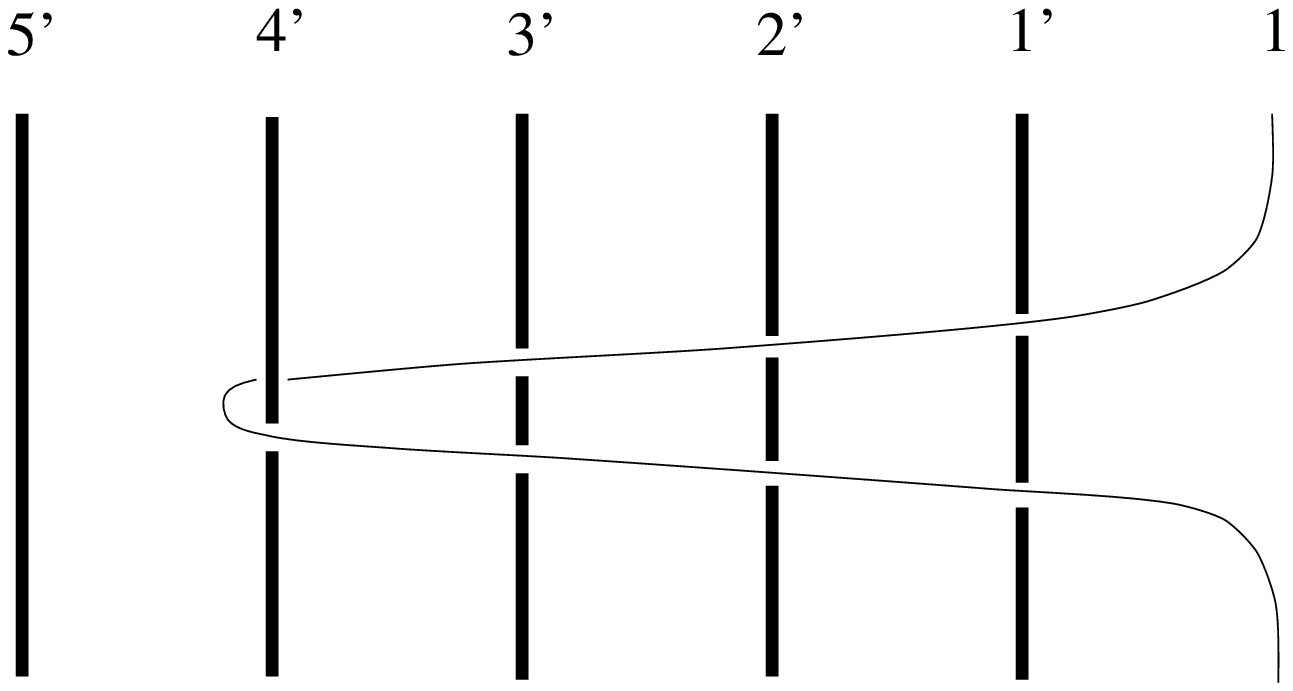,width=8cm}}
\centerline{Fig.7: \  The generator $\t_4$ for the $g=5$ case}}
\vskip0.5cm
\noindent
It is clear that $\t_\a$ acting on $V(g;n) = V_{g'} \otimes \cdots
V_{1'} \otimes V_1 \otimes \cdots V_n$ is represented as in \ident\
with the non-trivial part of $K_{(\a)}$ given by (identify indices
$0' \equiv 1$ where neccessary)
$$ \eqalign{K_{(\a)} &= q^{-1} \Rh^{-1}_{1'1} \Rh^{-1}_{2'1'} \cdots
\Rh^{-1}_{(\a-1)'(\a-2)'} \Rh^2_{{\a}'(\a-1)'}
\Rh^{\phantom{-1}}_{(\a-1)'(\a-2)'} \cdots \Rh^{\phantom{-1}}_{2'1'}
\Rh^{\phantom{-1}}_{1'1} \cr
&= \rho_f(\k+{\a})\ ,}   \eqn \krepgen  $$
and $\k+{\a}$ is expressed in terms of quantum algebra generators as
in \kpmm. Note that numbering the primed spaces is by convention, and
we have chosen a more natural one with opposite ordering compared to
\kpmm\ where it is fixed (cf. remark at end of section 2). All that
was said about the $g=2$ case above can be generalized to arbitrary
genus, there are no new features emerging.

We finish this section by noting that the generators $\t_\a$ in the
representation \krepgen\ can be considered as a subgroup of the
coloured braid group $C_{g+1}$. By definition, the coloured braid group
$C_{g+1}$ is the kernel of the mapping from the Artin braid group
$B_{g+1}$ to the permutation group $P_{g+1}$ having
${1 \over 2}g(g+1)$ elements $\kappa^{\phantom{-1}}_{\a \b}$ that can
be taken in our notation as
$$ \kappa_{\a \b} = \s^{-1}_{(\b+1)'} \cdots
\s^{-1}_{(\a-1)'} \  \s^{2 \phantom{-}}_{\a'}
\s^{\phantom{-1}}_{(\a-1)'} \cdots
\s^{\phantom{-1}}_{(\b+1)'} \ , \qquad 0 \leq \b < \a \leq g
\eqn \colbr $$
acting on $V(g;1) = V_{g'} \otimes \cdots V_{1'} \otimes V_1$. This
corresponds to $g$ bars plus one strand. Of course, in our case the
bars cannot wind around each other (but see the interpretation of
bars as lines in the next section) so we have to fix $\b = 0$ and let
$1 \leq \a \leq g$, s.t. $\kappa^{\phantom{1}}_{\a 0}$ gives
precisely the $g$ generators in \krepgen\ as a subset of the
generators of $C_{g+1}$. Therefore we can also think of $B^g_n$ as a
subgroup of the braid group $B_{g+n}$ having generators
$ \s^{\phantom{1}}_{g'}, \ldots, \s^{\phantom{1}}_{1'},
\s^{\phantom{1}}_1, \ldots, \s^{\phantom{1}}_{n-1}$. From our
experiences in section 2 it is obvious how to represent the full
coloured braid group in terms of quantum algebra generators, namely
the equivalent of \colbr\ is given by
$$ \eqalign{\k+{j;m} = \h+{g-j} \cdots \h+{g-j-(m-2)}
K^{\phantom +}_{g-j-(m-1)} \l+{g-j-(m-2)} \cdots \l+{g-j} , \qquad
j &= 0,\ldots,g-1  \cr  m &= 1,\ldots,g-j} \eqn \fcolbr $$
and similarly for $\k-{j;m}$ which gives \colbr\ but with all braid
group generators except $\s^2_{\a'}$ replaced by their inverses. For
each fixed value of $j$ they have commutation relations \recomgen.
Of course, we can write down a similar formula for $\k0{j;m}$ but
they do not represent the coloured braid group. It was already noted
in \refmark{\rSWZ} that the coloured braid group has a representation
within tensor products of the universal enveloping algebra of \sl{N}.


\chapter{INVARIANTS OF LINKS ON HANDLEBODIES}

We will now define links on the handlebody $H_g$ and then try to find
invariant polynomials. A $g$-link $L_g$ on $H_g$ is obtained as the
closure of a $g$-braid by connecting $P^{(0)}_i$ with $P^{(1)}_i$
outside the unit cube in the $x > 0 $ region (Fig.1). Citing a
theorem \refmark{\rSos}, every $g$-link can be obtained as the
closure of a $g$-braid. Markov moves for $B_n^g$ are defined
in the same manner as the usual ones for $B_n$, i.e. $B \rightarrow
B' B (B')^{-1}$ for arbitrary $B' \in B^g_n$ (Markov I) and $ B
\rightarrow B \s_n^{\pm 1}$ with $\s_n \in B^g_{n+1}$ (Markov II).
Then the Markov theorem would state that two $g$-braids have
equivalent closures iff there is a finite sequence of Markov moves of
type I and II taking one $g$-braid to the other. However, the Markov
theorem for $g >1$ was only stated as a conjecture in \refmark{\rSos},
it holds for $g = 1$ (we shall not need it in what follows).

There are several approaches to the construction of link polynomials,
one may roughly distinguish them in the following way (a convenient
access to original literature is \Ref\rKoha{\Koha}, basic accounts of
knot theory and the relation to quantum groups are e.g.
\REF\rGuaa{\Guaa}
\REF\rAGS{\AGS} \refmark{\rGuaa,\rAGS}). It is well known that the
expression of $\s_i$ in terms of $\Rh$ gives rise to a Hecke algebra
representation of the braid group $B_n$ and the characteristic
equation of the $\Rh$-matrix together with the first two equations of
\bralg\  comprise just the relations of the Hecke algebra $H(q^2,n)$
with generators $\s_i$ (we would have to rescale $q \rightarrow
q^{1/2}$ to make contact with the usual convention). One defines a
linear functional on $H(q^2,n)$, the Ocneanu trace, which is the main
ingredient in the definition of the invariant link polynomial
\Ref\rJon{\Jon}. This is just the quantum trace of the braid
group generators represented by $\Rh$ and the last step is then
proving invariance of it w.r.t Markov moves \REF\rTur{\Tur}
\REF\rReTu{\ReTu} \REF\rADW{\ADW} \refmark{\rTur,\rReTu,\rADW}.
Further it is possible to define link polynomials recursively using
skein relations \REF\rAlex{\Alex} \REF\rCon{\Con} \REF\rKauf{\Kauf}
\refmark{\rAlex,\rCon,\rKauf}. Finally, there is the Chern-Simons
field theory approach \REF\rWita{\Wita} \REF\rGMM{\GMM}
\refmark{\rWita,\rGMM}. In view of this we might expect that the
explicit representation \ident\ can be used to define an invariant
link polynomial on $H_g$ by means of quantum traces of generators
$\s_i$ and $\t_\a$. Also their characteristic equations should give
rise to skein relations.

We recall the definition of the Jones polynomial \refmark{\rJon}
$$ V(B) = q^{-3w({\hat B})} Tr \big\vert_{V(n)} (B \mu^{\otimes n}) ,
\eqn \jopol $$
which is a class function on the braid group $B_n$ that is invariant
w.r.t Markov moves of type I and II. Because of inclusion of type II
moves it is an ambient isotopic invariant, i.e. locks in a line are
irrelvant. In \jopol\ we denoted the closure of the braid $B \in B_n$
by ${\hat B}$, and $w({\hat B})$, the writhe or Tait number, is
the number of overcrossings minus the number of undercrossings in
${\hat B}$. Finally, the matrix $\mu$ is defined via an element of the
quantum algebra
$$ \mu = \pmatrix{q^{-1}&0 \cr 0&q} \equiv \rho_f(q^{-H}) , \eqn \mdef
$$
so the trace in \jopol\ could equally well be called the quantum trace
of $B$. For instance, the quantum trace of the \rea\ can be expressed
as $c_1=Tr(K\mu)$. Invariance of \jopol\ under Markov I rests on the
property $\lbrack \s_i, \mu^{\otimes n}\rbrack = 0$, and for Markov II
relies on the key property of the quantum trace $ Tr\big\vert_{V_{n+1}}
(\s^{\pm 1}_n (I^{\otimes n}\otimes \mu)) = q^{\pm 3} I^{\otimes n}$.
The Jones polynomial satisfies a skein relation which is obtained from
the characteristic equation of $\s_i$
$$ q^{-1} \s_i - q \s_i^{-1} - \w I = 0  \eqn \schar $$
by using linearity of the trace, the result is for arbitrary $B,B' \in
B_n$
$$ q^2 V(B \s_i B') - q^{-2} V(B \s_i^{-1} B') - (q-q^{-1}) V(BB') = 0
. \eqn \jposk $$
This can be depicted as
\vskip1cm \vbox{
\centerline{\psfig{figure=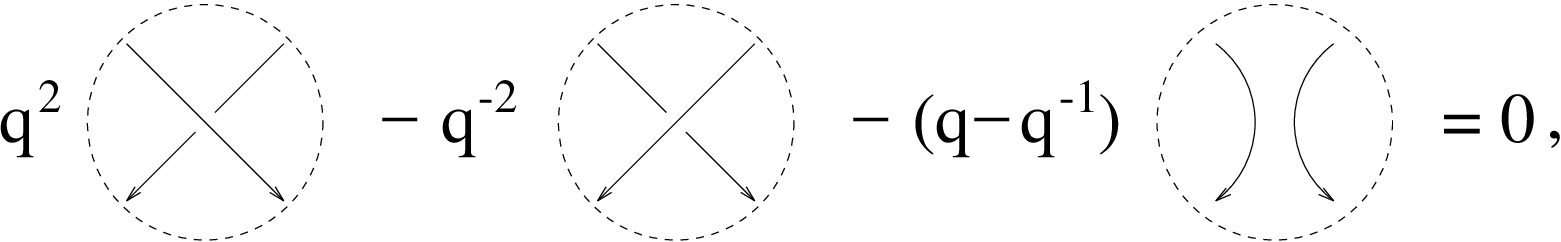,width=12cm}}
\centerline{Fig.8: \  Skein relation of the Jones polynomial}}
\vskip0.5cm \noindent
where each pictogram means the polynomial of the closed braid
differing only at the crossing between $B$ and $B'$ as indicated.
By simply closing the lines the normalization of the unknot is
obtained as $N=q+q^{-1}$, this is the same as in \refmark{\rWita} up
to rescaling of $q$ and corresponds to standard framing (see the
discussion in \Ref\rWitb{\Witb} on the effects of framing).

There is a simple possibility to make use of this formalism in our
context as we have a representation of $\t_\a$ in terms of
$\Rh$-matrices. We just extend the definition of the Jones polynomial
from $V(n)$ to $V(g;n)$ and get an ambient isotopy invariant
polynomial for $g$-links on $H_g$ in terms of the Jones polynomial of
links in ${\bf {\rm R}}^3$ (or a 3-ball, for that matter). Proof of
invariance w.r.t. Markov I,II works as before. We are forced to
interprete bars corresponding to the handles as lines and, because of
the trace in $V(B)$, close them to obtain the link in ${\bf {\rm
R}}^3$ which is associated to the $g$-link on $H_g$.  Because we have
an description of the crossings by $\Rh$-matrices the direction of the
bars is fixed downwards. This procedure is obviously well defined as
the equivalence class of a $g$-link is mapped uniquely onto the class
of the associated ordinary link \refmark{\rSos}. The polynomial for
$g$-links is then given by
$$ V_g(B) = q^{-3w({\hat B})} Tr \big\vert_{V(g;n)} (B \mu^{\otimes
(g+n)}) , \eqn \gjopol $$
where $B \in B_n^g$ is a word in the generators $\s_i$ and
$\t_\a$, its closure is obtained via the representation \ident\ and
\krepgen\ of $\t_\a$. It is easy to find examples of $g$-links which
belong to different classes but share the same value of the
polynomial.

\noindent
It might appear as if one were back to the usual situation where one
deals with the generators $\s_i$ only. However, keeping generators
$\t_\a$ is a great advantage, the reason is that they are very special
expressions in $\s_i$. They all obey the characteristic equation
$$ q^{-2} \t_\a + q^2 \t_a^{-1} - c_1 I = 0 , \eqn \tchar $$
which, as before, leads to a skein relation that follows from \gjopol
$$  q^4 V_g(B \t_\a B') + q^{-4} V_g(B \t_\a^{-1} B') - (q^2 + q^{-2})
V_g(BB') = 0  \eqn \gjposk $$
in addition to the $\s_i$ skein relation for $V_g(B)$. Here we
inserted the value $c_1 = q^2 + q^{-2}$ for the
fundamental representation of \sl2. Above relation can be depicted as
\vskip1cm \vbox{
\centerline{\psfig{figure=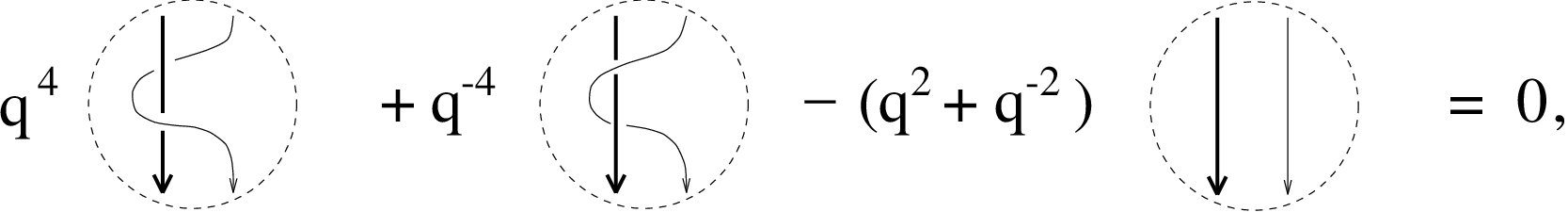,width=14cm}}
\centerline{Fig.9: \  The additional skein relation for generators
$\t_\a$}}
\vskip0.4cm \noindent
where we displayed only the $g=1$ case. It is known
\refmark{\rAlex,\rCon} that by recursively using skein relations the
invariant polynomial can be calculated uniquely. So it is reasonable
to suggest that both skein relations \jposk\ and \gjposk\ suffice to
calculate a well defined polynomial for any $g$-link \refmark{\rSch},
knowing the origin of the additional skein relation it is obvious
that this statement is correct.

If we use skein relations to calculate the polynomial we can fix the
procedure as follows. First untie the knot in the topological trivial
region using \jposk, this is clearly possible and it eventually gives
unknots going around the bars. If an unknot winds around a bar $n$
times, $(\t_\a)^n$, then it always can be reduced to $n=1$ with the
help of \gjposk, regardless whether $n$ is positive or negative.
Similarly if it winds around several bars in the right order by using
the analogue \comcen\ of \gjposk\ for a product of several generators,
otherwise the correct order must be established first by using \gjposk.
This way the reduction process can be much simplified, but
nevertheless in the end \jposk\ has to be used again to untie the
simple loops around the $S^1$-factors (closed bars). If there is only
one loop simple rules can be established as indicated in Fig.10a.
\vskip1cm \vbox{
\centerline{\psfig{figure=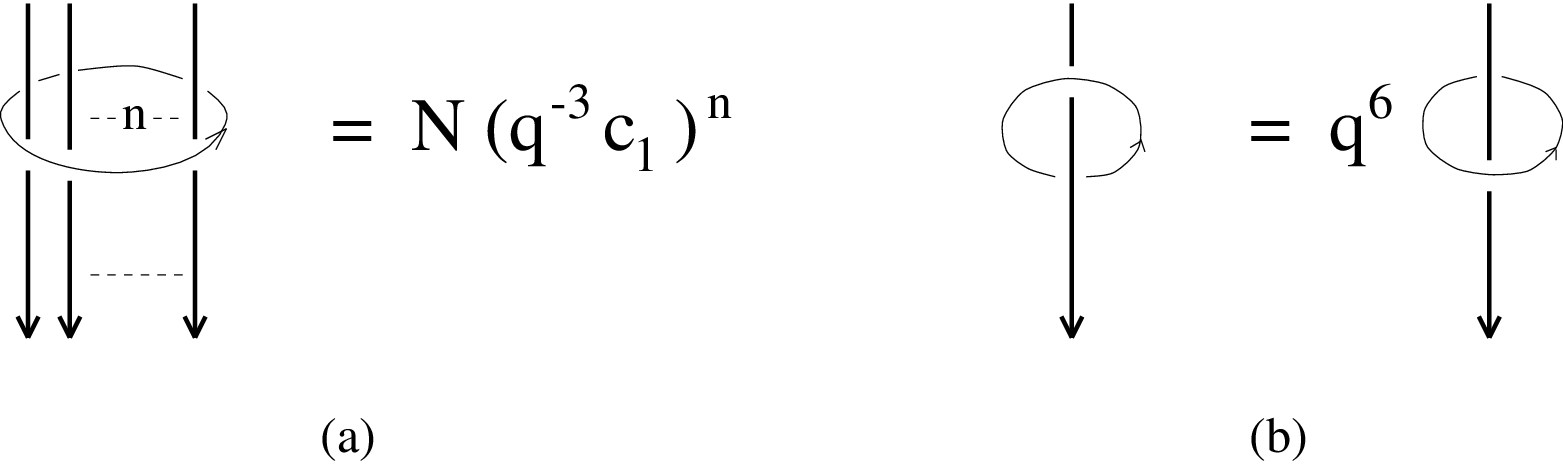,width=12cm}}
\centerline{Fig.10: \  (a) Evaluation of a simple loop encircling $n$
(closed) bars ($N=q+q^{-1}$ is the}
\centerline{normalization of the unknot), (b) Relation between
$\t_\a^{-1}$ and $\t_\a$ {\phantom {xxxx}}}}
\vskip0.4cm \noindent
The virtue of Fig.10a is that it connects loops going around handles
to loops in topologically trivial regions, we can just reinsert
instead of the factor $N$ an unknot in ${\bf {\rm R}}^3$ (if there are
bars not being encircled by the loop they contribute factors of $N$ on
the RHS). In a way, it looks as if this were related to the surgery
method described in \refmark{\rWita}, see also \REF\rGuab{\Guab}
\REF\rGuac{\Guac} \refmark{\rGuab,\rGuac}.
One can also express a loop originating from $\t_\a^{-1}$ directly in
terms of one originating from $\t_\a$ (only the $g=1$ case is shown in
Fig.10b), if the loop encircles $n$ closed bars corresponding to a
ordered product of $n$ generators $\t_\a$ the factor will be $q^{6n}$.
This can be taken a little further but as we do not have concrete
calculations in mind we do not elaborate on it.

Of course, even though it is possible to define the above invariant
polynomial, we would have prefered to define it intrinsically on the
handlebody keeping the information about the topology strictly, i.e.
without transforming holes into lines carrying some representations.
But this is not so easy, because if we use \jopol\ defined on $V(n)$
only, but with $g$-braids $B$ containing generators $\t_\a$, then the
trace is no longer invariant with respect to Markov I. This follows
from the fact that $\lbrack K,\mu \rbrack \not= 0$, the only
difference occurs in first space and Markov II is still valid. We are
presently looking for a modification of the polynomial that would
employ only the algebraic properties of the \rea\ and make no use of
the representation discussed above, whether this is possible and
whether it would lead to an inequivalent invariant is an open 
question.


\chapter{DISCUSSION}

There are a few topics that can be mentioned in connection with the
present work. The motivation in \refmark{\rSos} was to define
invariant polynomials of links intrinsically on any closed 3-manifold
$M$. The prerequisite for this is a polynomial defined on
handlebodies, it would then be neccessary to investigate how it
transforms w.r.t. the Heegard homeomorphism $\psi : \partial H_g
\rightarrow \partial H_g$ since any closed compact 3-manifold $M$ can
be obtained by the Heegard decomposition $M=H_g \cup_\psi H'_g, \ H_g
\cap H'_g = \partial H_g = \partial H'_g$.  Every link in $M$ is
isotopic to a closed braid in $H_g$, but the braid depends on the
Heegard splitting. One would then need to use (a subgroup of) the
mapping class group of a genus $g$ Riemann surface in order to study
the behaviour of the polynomial w.r.t. the Heegard homeomorphism,
maybe the approach in \Ref\rKohb{\Kohb} could be useful where the
homeomorphisms of a handlebody were expressed essentially in terms of
$R$-matrices.  Related to this subject is the surgery method because
the Heegard splitting could be used also to transform invariants that
are defined on a certain closed 3-manifold to a different one. Whether
such an approach would be more tractable compared to \refmark{\rWita}
is not clear a priori. The presence of the generators $\t_\a$ suggests
the idea of an `algebraization' of the surgery method.

A further problem is whether there exist representations of $B^g_n$
other than in terms of $R$-matrices. The \rea\ as written in \alg,
especially existence of central elements and characteristic equations,
are a consequence of the Hecke algebra representation of the
generators $\s_i$. It is not clear whether it is possible to represent
$\t_\a$ differently from $\s_i$. The problem to find representations
of $B^g_n$ is of course related to the difficulty in definig invariant
link polynomials, because the Markov trace is dependent on classes of
(irreducible) representations. In this context it is worthwhile noting
that at least $B^1_n$ is a Coxeter group \refmark{\rSos}.

Also, we would like to draw attention again to the (infinite
dimensional) extended \rea\ and the new representations of it that we
constructed in terms of quantum algebra generators which satisfy the
system of commutation relations \recomgen. This might have some
applications other than discussed here, e.g. in the description of
differential geometry on quantum groups or quantum spin chains. 
During typesetting this manuscript we came across two preprints which
bear some similarity to our work. In \Ref\rAle{\Ale} the monodromies
of flat connections around the cycles of a Riemann surface with marked
points are considered, they obey some variant of the extended \rea.
In \Ref\rKaZa{\KaZa} a quantum group invariant $n$-state vertex model
on a torus is constructed which has a topological interaction of the
vertices with the interior of the torus, the grahical notation is also
reminiscent of ours.

\vfill
\eject
\refout
\bye